\documentclass[conference]{IEEEtran}
\IEEEoverridecommandlockouts
\usepackage{cite}
\usepackage{amsmath,amssymb,amsfonts}
\usepackage{algorithmic}
\usepackage{graphicx}
\usepackage{textcomp}
\usepackage{xcolor}
\def\BibTeX{{\rm B\kern-.05em{\sc i\kern-.025em b}\kern-.08em
    T\kern-.1667em\lower.7ex\hbox{E}\kern-.125emX}}

\usepackage{verbatim}
\usepackage{comment}

\graphicspath{ {./image/} }

\def \cC{\mathcal{C}}

\makeatletter
\newcommand\fs@spaceruled{\def\@fs@cfont{\bfseries}\let\@fs@capt\floatc@ruled
  \def\@fs@pre{\vspace{1\baselineskip}\hrule height.8pt depth0pt \kern2pt}%
  \def\@fs@post{\kern2pt\hrule\relax}%
  \def\@fs@mid{\kern2pt\hrule\kern2pt}%
  \let\@fs@iftopcapt\iftrue}
\makeatother

\begin{document}

\title{Keep the bursts and ditch the interleavers\\
}
\author{\IEEEauthorblockN{Wei An}
\IEEEauthorblockA{\textit{Research Laboratory of Electronics} \\
\textit{Massachusetts Institute of Technology}\\
Cambridge, MA 02139, USA \\
wei\_an@mit.edu}
\and
\IEEEauthorblockN{Muriel M\'edard}
\IEEEauthorblockA{\textit{Research Laboratory of Electronics} \\
\textit{Massachusetts Institute of Technology}\\
Cambridge, MA 02139, USA \\
medard@mit.edu}
\and
\IEEEauthorblockN{Ken R. Duffy}
\IEEEauthorblockA{\textit{Hamilton Institute} \\
\textit{Maynooth University}\\
Ireland \\
ken.duffy@mu.ie}
}

\maketitle

\begin{abstract}
To facilitate applications in IoT, 5G, and beyond, there is an engineering need to enable high-rate, low-latency communications. Errors in physical channels typically arrive in clumps, but most decoders are designed assuming that channels are memoryless. As a result, communication networks rely on interleaving over tens of thousands of bits so that channel conditions match decoder assumptions. Even for short high rate codes, awaiting sufficient data to interleave at the sender and de-interleave at the receiver is a significant source of unwanted latency. Using existing decoders with non-interleaved channels causes a degradation in block error rate performance owing to mismatch between the decoder's channel model and true channel behaviour.

Through further development of the recently proposed Guessing Random Additive Noise Decoding (GRAND) algorithm, which we call GRAND-MO for GRAND Markov Order, here we establish that by abandoning interleaving and embracing bursty noise, low-latency, short-code, high-rate communication is possible with block error rates that outperform their interleaved counterparts by a substantial margin. Moreover, while most decoders are twinned to a specific code-book structure, GRAND-MO can decode any code. Using this property, we establish that certain well-known structured codes are ill-suited for use in bursty channels, but Random Linear Codes (RLCs) are robust to correlated noise. This work suggests that the use of RLCs with GRAND-MO is a good candidate for applications requiring high throughput with low latency.
\end{abstract}


\section{Introduction} \label{sect:intro}

Raw communication channels are never memoryless. Channel fading, inter-symbol interference, multi-user interference, and external noise sources all have inherent time-scales that result in time-dependent correlations in instantaneous Signal to Interference plus Noise Ratio (SINR). Essentially all forward error correction decoders assume, however, that channels are memoryless \cite{Goldsmith2005,shu2004} and, as we shall demonstrate, their performance degrades significantly if they are not. 

\begin{figure}[htbp]
\centerline{\includegraphics[width=0.49\textwidth]{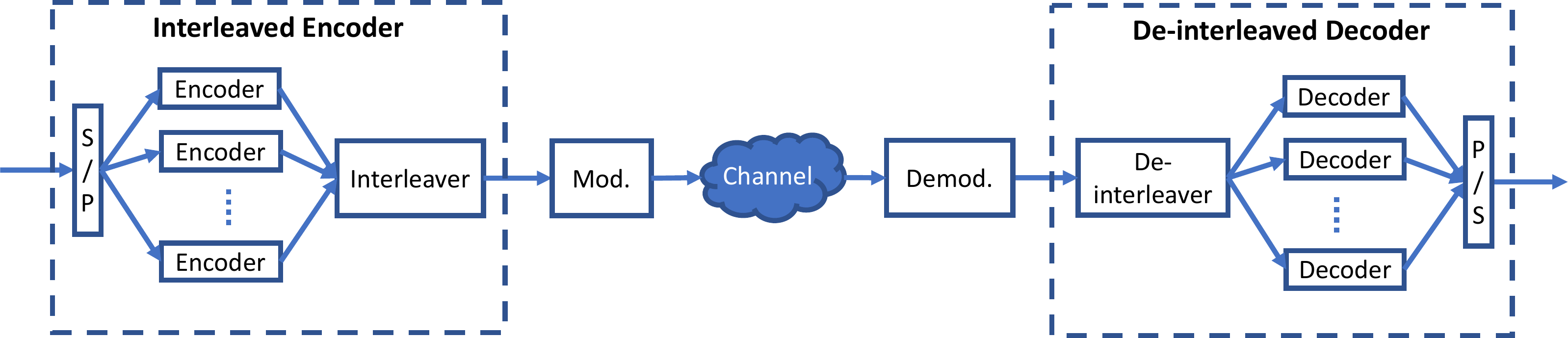}}
\caption{Interleavers in communication systems introduce significant latency.}
\label{fig:intl_diag}
\end{figure}

As illustrated in Fig. \ref{fig:intl_diag}, the engineering solution to this mismatch is to employ interleaving. The transmitter takes data, adds forward error correction, and buffers a large number of the resulting code-words. The interleaver permutes the location of bits across the code-words prior to their transmission. On the receiver side, the permutation of bits is first inverted by the de-interleaver, resorting the original bit order, and the resulting signals are passed to a decoder for error detection and correction. In this way, errors that are clumped in the transmitted bit order are separated and distributed across multiple code-words, giving noise on the medium the appearance of having less memory.

Even if short codes, such as the CA-Polar codes proposed for 5G-NR control communications \cite{Arikan09,KK-CA-Polar,TV-list}, are employed, owing to the requirement to permute bits across a large spacing, transmitters need to wait for a sufficient volume of data before interleaving, resulting in undesirable delays of the order of those experienced by large packet transmissions consisting of thousands of bits. There are many emerging applications, however, where this latency is particularly unwelcome. Examples include machine communications in IoT systems \cite{ieee_latency, iot2019, wHP_fec}, telemetry, tracking, and command in satellite communications \cite{ttc_2005, fec_ttc_2014}, all control channels in mobile communication systems, and Ultra-Reliable Low Latency Communications (URLLC) as proposed in 5G-NR \cite{urllc,sybis2016channel,shirvanimoghaddam2018short}. 

The only solution to this conundrum is to develop codes and decoders that can provide effective error correction in correlated noise channels without resorting to interleaving. By further developing the recently introduced Guessing Random Additive Noise Decoding (GRAND) \cite{Duffy18,kM:grand} algorithm to avail of the correlation in noise, here we propose one such solution.

While it has long-since been known that it is computationally infeasible to create a universal decoder that works with all arbitrary long codes \cite{berlekamp1978inherent,bruck1990hardness}, which necessitated the co-design of code-books and code-specific decoders \cite{lin2004error} for long codes, the same is not true for short codes. In particular, GRAND is a universal decoding scheme that can work efficiently with any short, high-rate code. GRAND's universality stems from its effort to identify the effect of the noise, from which the code-word is deduced.

The original design established mathematical properties of a class of Maximum Likelihood hard-detection decoders \cite{Duffy18,kM:grand}, but omitted important practical implementation details that we complete here for Markovian channels (to be described in section \ref{sub:Markov}). For soft detection with increasing levels of quantization and memoryless channels, a series of variants have been proposed: SRGRAND \cite{Duffy19a,Ken:5G}, ORBGRAND \cite{Duffy20} and SGRAND \cite{Solomon20}. In the current work, we adapt GRAND to make it suitable for channels that induce Markov correlated error bursts in a hard-detection setting, establishing that significant gains in Block Error Rate (BLER) performance are possible for short, high-rate code, obviating the need to use interleaving and thus enabling URLLC.

\begin{figure}[htbp]
\centerline{\includegraphics[width=0.49\textwidth]{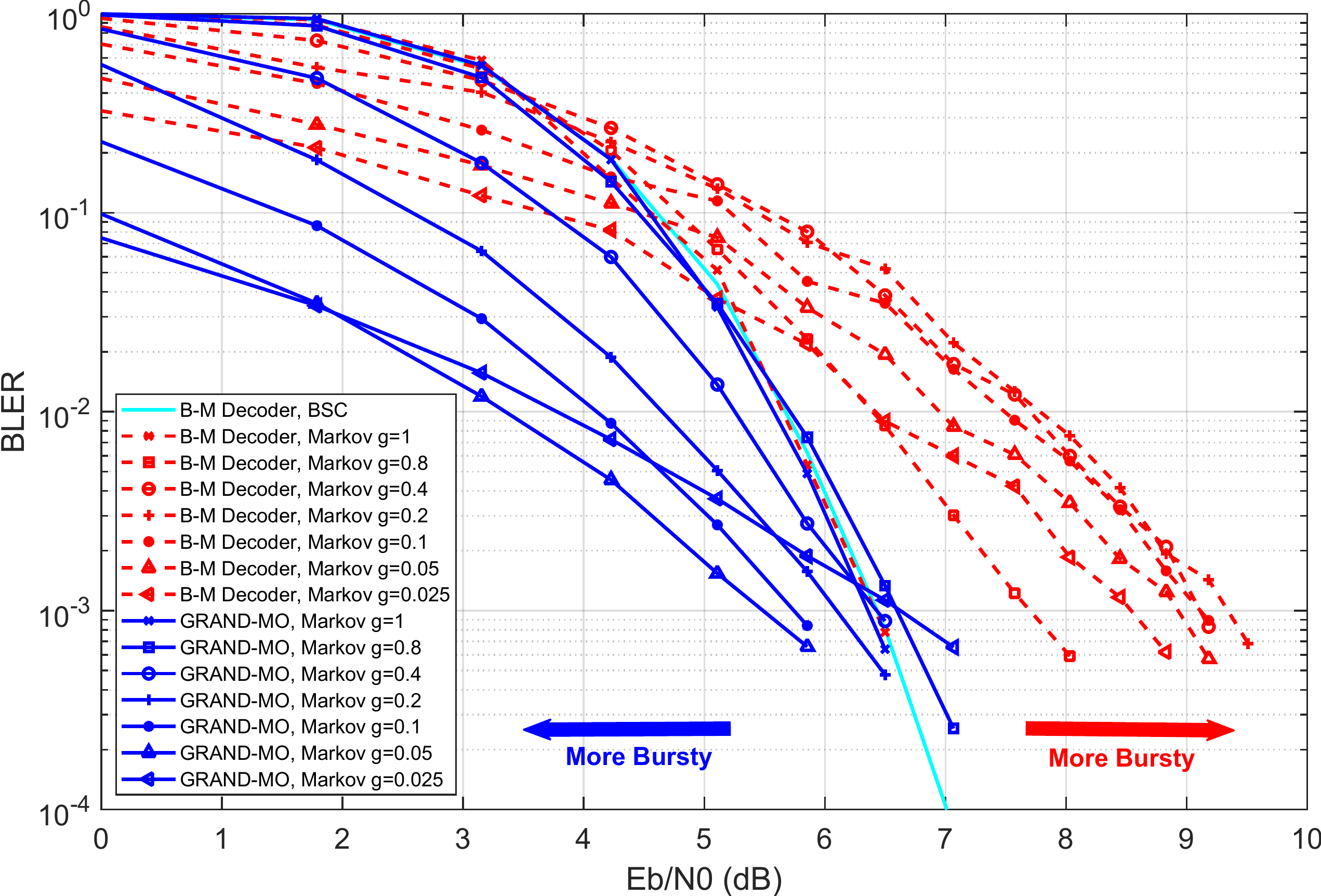}}
\caption{Performance of BCH(127,106) with the Berlekamp-Massey (B-M) and  GRAND-MO decoders in BSC and Markov channels.}
\label{fig:bch_bsc_markov_multi_burst}
\end{figure}
While GRAND-MO is a universal decoder, its BLER performance depends on the code book structure used and its code rate. In general, we observe that it works best in bursty channels with code books that have limited structure. Fig. \ref{fig:bch_bsc_markov_multi_burst} provides on example of the GRAND-MO as applied to a BCH(127,106) code in Markov channels where the average error burst length is $1/g$. GRAND-MO's decoding performance improves for channels with increasing memory, i.e. as $g$ decreases. In contrast, the performance of the standard BCH Berlekamp-Massey (B-M) decoder \cite{journals/iandc/BoseR60a} is also presented, which degrades quickly with increasing average error burst length. For highly correlated bursts, at a target BLER of $10^{-3}$ GRAND-MO sees a greater than 3dB gain when compared to the B-M decoder. This dramatic difference results from the fact that GRAND-MO can exploit the structure of error bursts, instead of dumping them, as standard decoders require. 

Notably, even with GRAND-MO, the BCH(127,106) code loses its robustness in a channel where the lengths of the bursts are comparable to that of the code-word, e.g. $g=0.025$, so that the average error burst length is $40$ bits. This arises because the structure of BCH code-words clashes with typical burst patterns. Using GRAND-MO's universality, later we shall demonstrate that Random Linear Codes (RLCs) do not possess this shortcoming.

The rest of the paper is organized as follows. Section \ref{sect:markov} introduces GRAND and the Markov channel model. Section \ref{sect:pattern} provides an algorithmic description of our GRAND-MO error pattern generator, which  matches the Markov channel. Section \ref{sect:app_sim} provides simulated performance evaluation for Reed-Muller, BCH  and RLC codes, and assesses the impact of interleavers on performance. Section \ref{sect:conclusion} summarizes the paper.

\section{GRAND for Markov Channels} \label{sect:markov}

\subsection{Guessing Random Additive Noise Decoding} \label{sub:grand}
Unlike most deployed decoders, GRAND focuses on identifying the noise that impacted a communication rather than the code-word itself. Consider a transmitted binary code-word $X^n\in {\cC}$ draw from an arbitrary rate $R$ code-book  ${\cC}$, i.e. a set of $2^{nR}$ strings in $\{0,1\}^n$. Assume independent channel noise, $N^n$, which also takes values from $\{0,1\}^n$, additively alters $X^n$ between transmission and reception. The resulting sequence is $Y^n = X^n \oplus N^n$, where $\oplus$ represents addition modulo 2.

From $Y^n$, GRAND attempts to determine $X^n$ indirectly by identifying $N^n$ through sequentially taking putative noise sequences, $z^n$, which we sometimes term patterns, subtracting them from the received signal and querying if what remains, $Y^n\oplus z^n$, is in the code-book $\cC$. If transmitted code-words are all equally likely and $z^n$ are queried in order from most likely to least likely based on the true channel statistics, the first instance where a code-book element is found is an optimal maximum likelihood decoding \cite{kM:grand}. GRAND's premise is that for short, high-rate codes, there are much fewer potential noise sequences than code-words. In addition, it is established in \cite{kM:grand} that one need not query all possible noise patterns for the decoder to be capacity-achieving, and instead can determine a threshold number of queries, termed abandonment threshold, at which a failure to decode can be reported without unduly impacting error correction performance. 

Central to GRAND's theoretical performance is the querying of putative noise sequences with a likelihood order that matches the channel statistics. Core to its practical realisation is the ability to do so efficiently, but that is not described in the original paper \cite{kM:grand}; for Markov channels, this is what GRAND-MO achieves.

\subsection{Markov Channel and GRAND-MO} \label{sub:Markov}
We consider a classic two-state Markov chain to model a binary channel with bursty errors \cite{Gilbert1960}. When in the good state, $G$, the channel is error-free and the corresponding entry in $N^n$ is to a $0$, and whenever the channel is in the bad state, $B$, there is an error and the corresponding entry in $N^n$ is to a $1$. The transition probability from $G$ to $B$ is $b$ and from $B$ to $G$ is $g$. An error burst is a set of consecutive 1s in $N^n$ with its length following geometry distribution of mean $1/g$ and variance $(1-g)/g^2$. Same to the length of an error-free run, a set of consecutive 0s in $N^n$, with mean of $1/b$. The stationary bit-flip probability of the Markov channel is $p = b/(b+g)$ and its correlation is $1-b-g$. Note that if $b=1-g$, the Markov channel is a memoryless BSC with $p=b$. Both $b$ and $g$ are assumed known and can  be estimated in practice.

We seek to understand the likelihood of error sequences so that in Section \ref{sect:pattern} we can identify  an algorithm that efficiently produces the $z^n$ patterns sequentially from most likely to least likely for use with GRAND. Consider that $N^n$ has $m$ bursts of errors, with $l$ total flipped bits. Three cases arise:
\begin{itemize}[
    \setlength{\IEEElabelindent}{\dimexpr-\labelwidth-\labelsep}
    \setlength{\itemindent}{\dimexpr\labelwidth+\labelsep}
    \setlength{\listparindent}{\parindent}
  ]
    \item Case 0: $N^n$ begins and ends with 0s. The probability of this case is
        \begin{equation} \label{eq:P_0}
            P_0(m, l) = \frac{g}{1-b}  \frac{(1-b)^n}{b+g}  \left( \frac{bg}{(1-b)(1-g)} \right) ^m \left( \frac{1-g}{1-b} \right) ^l.
        \end{equation}
    \item 
    Case 1: $N^n$ either begins or ends with a 1. The probability of this event is
    \begin{equation} \label{eq:P_1}
    P_1(m, l)= \frac{1-b}{g} P_0(m, l) 
    \end{equation}
    \item
    Case 2:  $N^n$ begins and ends with 1s, which occurs with probability
     \begin{equation} \label{eq:P_2}
     P_2(m, l)=\left(\frac{1-b}{g}\right)^2 P_0(m, l)
      \end{equation}
\end{itemize}

In all operating regimes, $p=b/(b+g)<1/2$ and we therefore assume that $b<g$. If the occurrence of errors is positively correlated, $0< 1-b-g$, so that $1< (1-b)/g$ and hence 
\begin{equation} \label{eq:P2P1P0}
P_0(m, l) < P_1(m, l) < P_2(m, l),
\end{equation}
while the reverse is true if the chain is negatively correlated. For fixed $m$, the probability decreases as $l$ increases as $b<g$. If there is no memory so that the channel is a BSC, i.e. $0=1-b-g$, it can be directly confirmed that $P_0(m, l) = P_1(m, l) = P_2(m, l) = (1-p)^{n-l}p^l$ and that $m$, the number of bursts, has no relevance.

\section{GRAND-MO Pattern Generator}  \label{sect:pattern}

\begin{figure}[htbp]
\centerline{\includegraphics[width=0.5\textwidth]{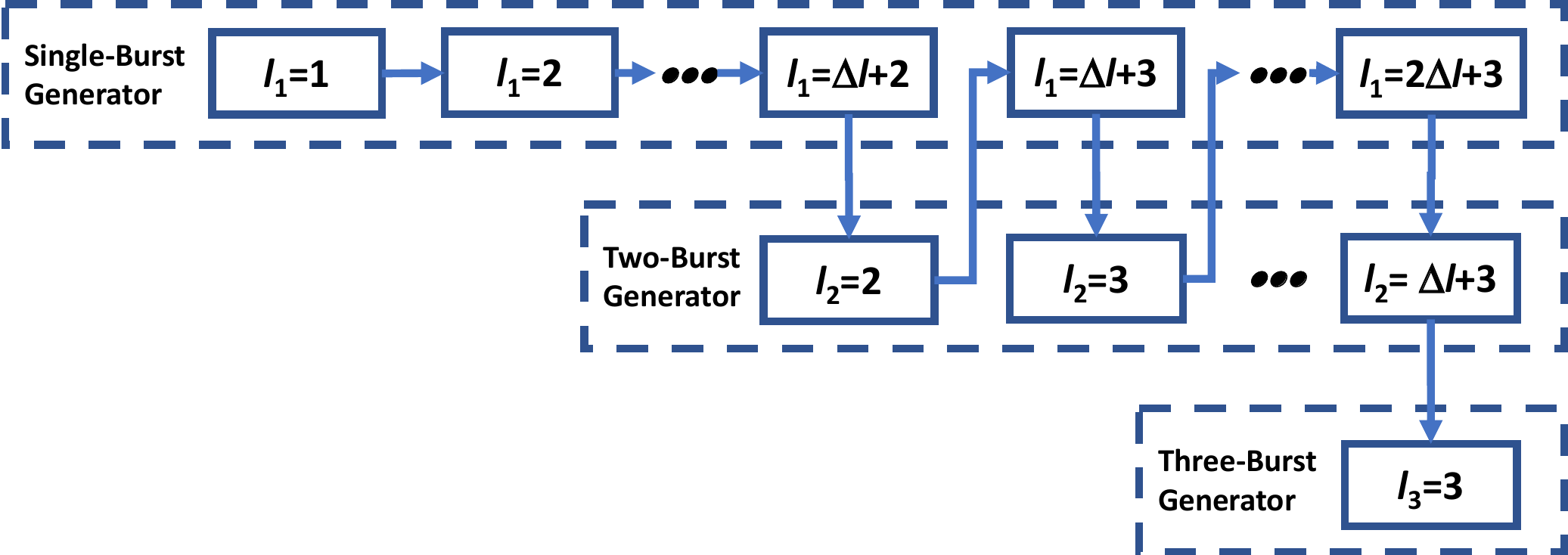}}
\caption{Illustration of Markov noise sequence generator}
\label{fig:mo_pat_gen}
\end{figure}

In order to complete the ordering we need to understand when, for a current $m$ and $l_m$, the next most likely pattern has the same $m$ and $l_m\mapsto l_m+1$ or we need to interlace with a pattern having an extra burst, $m\mapsto m+1$ with $l_{m+1}\geq m+1$. For bursty channels the chain will be positively correlated and so we make this determination by setting aaaaaaaaaaaa $P_0(m, l_m) = P_2(m+1, l_{m+1})$, from which we obtain
\begin{equation} \label{eq:deltaLb2}
    \Delta l = l_m -l_{m+1} =  \frac{\log \left (\frac{b}{g}\right )}{\log \left ( \frac{1-g}{1-b}\right )} -1 , 0 \leq \Delta l \leq n.
\end{equation}
When patterns with $m$ bursts contain $l_m > \Delta l + m+1$ flipped bits, patterns with $m+1$ bursts are interlaced into the generative order, starting with an initial $l_{m+1}=m+1$ flipped bits. For clarity, we illustrate the operation of the pattern generation algorithm in Fig. \ref{fig:mo_pat_gen}.

\begin{figure}[htbp]
\centerline{\includegraphics[width=0.49\textwidth]{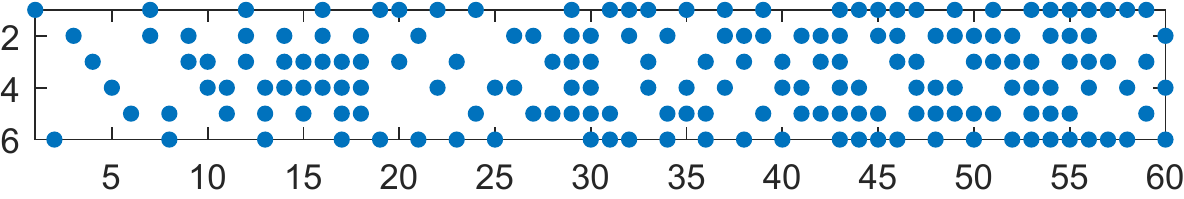}}
\caption{Order of patterns generated for use in GRAND-MO for a code word of length $n= 6$, $\Delta l=2$, maximum $m=3$ bursts, and last number of flipped bits  $l=3$. Columns indicate patterns, which are ordered left to right in decreasing likelihood.}
\label{fig:mo_pat}
\end{figure}

\begin{figure}[htbp]
\centerline{\includegraphics[width=0.4\textwidth]{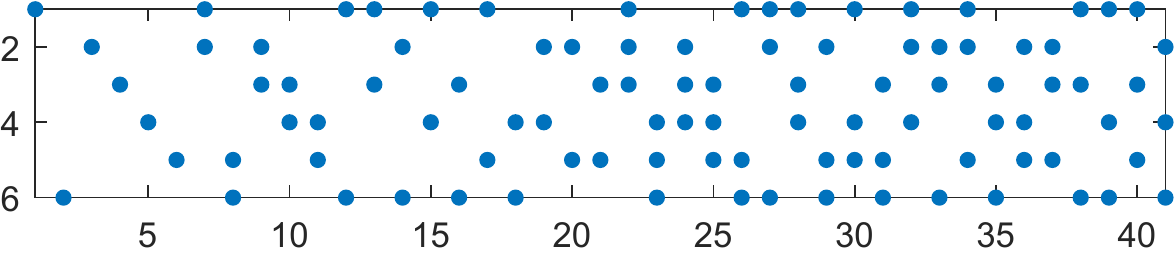}}
\caption{GRAND-MO pattern order for BSC channel, corresponding to $\Delta l=0$, for a code word of length $n=6$, maximum $m=3$ bursts, and final number of flipped bits $l=3$.}
\label{fig:sos_pat}
\end{figure}

Fig. \ref{fig:mo_pat} provides a visualization of the Markov query order for strings of length $n= 6$ with $\Delta l=2$ where each column corresponds to a putative noise sequence, a dot corresponds to a flipped bit, i.e. a 1, and the absence of a dot corresponds to a 0. Sequences are ordered in terms of likelihood from left to right. If the channel was memoryless, i.e. $\Delta l = 0$), these sequences would have non-decreasing numbers of bit flipped from left to right, as illustrated in Fig. \ref{fig:sos_pat}. In a Markov channel, however, due to the bursty structure of the noise, note that sometimes sequences of higher Hamming weight are queried before those with less. This important feature explains why the minimum Hamming distance of a code, which is an essential metric in a BSC, has significantly less relevance for bursty channels.

For a query abandonment criterion, let the maximum burst number be $m_{max}$ and the number of flipped bits when $m=m_{max}$ be $l_{last}$, then we abandon querying and report a decoding error when 
\begin{equation} \label{eq:ABml}
    m_{max} = l_{last} = \left \lfloor d/2 \right \rfloor
\end{equation}
where $d$ is the minimum Hamming weight of the codewords in ${\cC}$. With this abandonment condition, one may verify that the sequences checked by GRAND-MO for Markov channels with $\Delta l > 0$ is a super-set of the set of testing patterns for a BSC channels ($\Delta l=0$) with the same abandonment conditions, as can be seen in Fig. \ref{fig:mo_pat} and Fig. \ref{fig:sos_pat}. Since BSC is the channel that interleavers seek to emulate, such a guarantee is important in the context of this work. 

\begin{figure}[htbp]
\centerline{\includegraphics[width=0.5\textwidth]{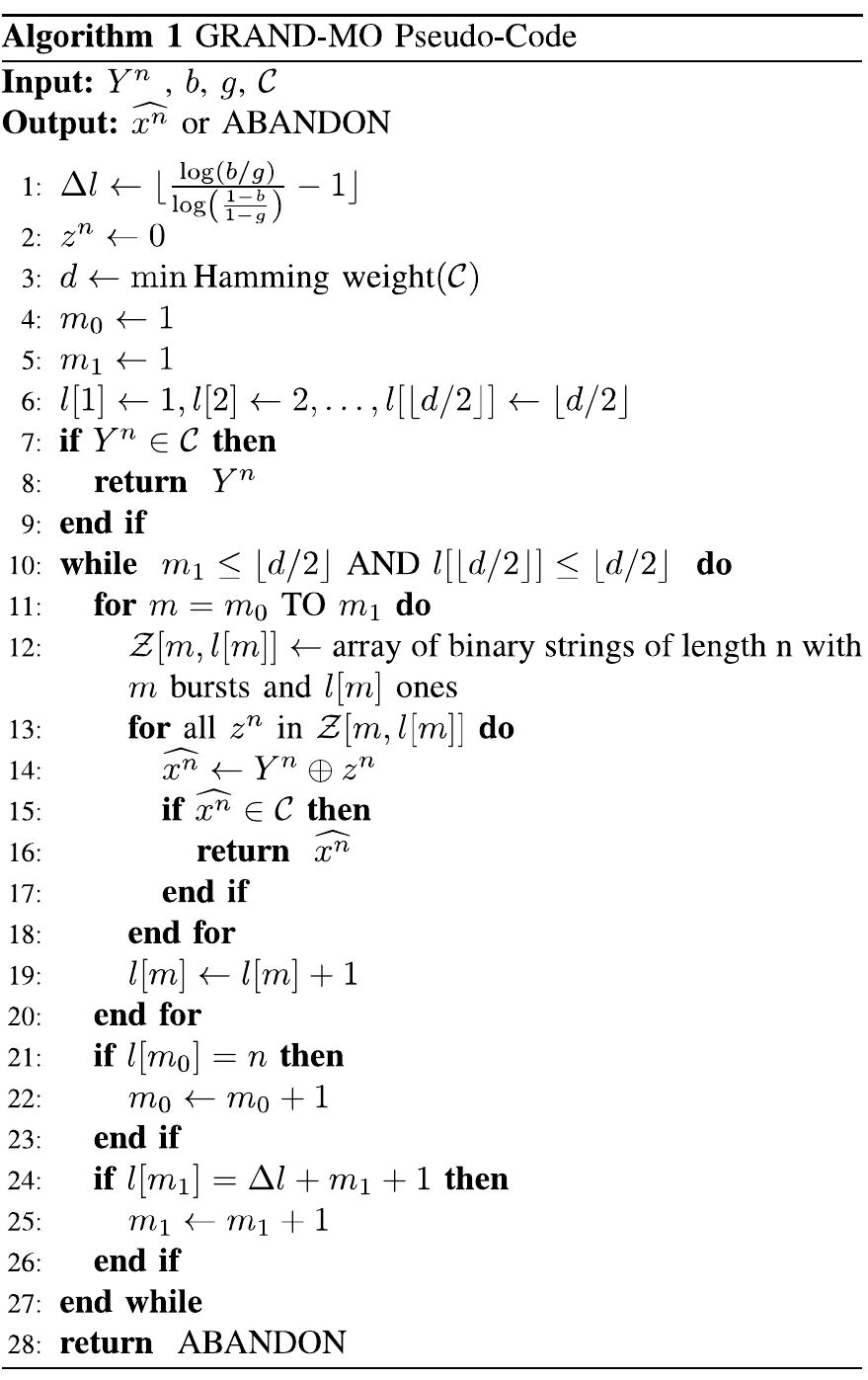}}
\end{figure}

Finally, having determined the appropriate order in which to query Markov noise sequences in decreasing probability with respect to $N^n$, pseudo code for GRAND-MO is presented in Algorithm 1. The decoder terminates when either $Y^n \oplus z^n$ is in the codebook or the abandonment criterion is reached.


Generally $\Delta l$ increases with channel memory via (\ref{eq:deltaLb2}). For a channel that is memoryless, $\Delta l=0$, corresponding to a BSC.

\section{Applications and Simulations} \label{sect:app_sim}

In the simulations, we map the stationary bit flip probability of the channel energy per transmitted information bit via the usual AWGN BPSK formula:
\begin{align*}
p= \frac{b}{b+g} = \text{erfc}\left(\sqrt{2RE_b/N_0}\right),
 \end{align*}
where $R$ is the coding rate.

\subsection{Comparison with standard decoders}
\label{subsect:standard}

\begin{figure}[htbp]
\centerline{\includegraphics[width=0.49\textwidth]{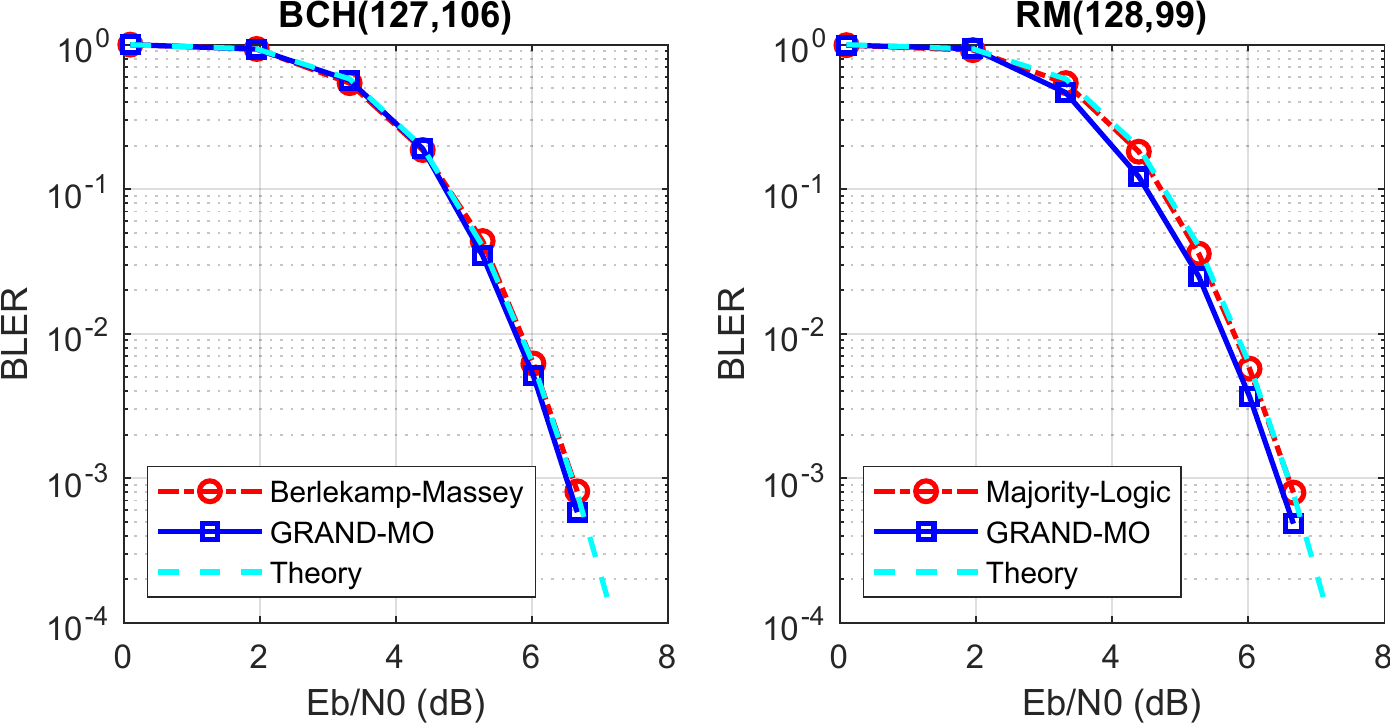}}
\caption{Performance of GRAND-MO in an idealized BSC channel compared to standard decoders, and with reference to theoretical random-error-correcting capability}
\label{fig:sos_bsc}
\end{figure}

We first evaluate the performance of GRAND-MO in a memoryless BSC setting corresponding to idealized infinite interleaving, i.e. $b=1-g$. We use two distinct, standard code-book constructions, BCH and Reed-Muller (RM), for which there exists well-established BSC hard detection algorithms: Berlekamp-Massey for BCH and Majority-Logic  \cite{journals/tc/Muller54} for RM. Fig. \ref{fig:sos_bsc} reports their performance along with standard theoretical bounds and GRAND-MO with an abandonment condition of $l_{max}=4$. Note that the random-error-correcting capabilities of Berlekamp-Massey decoding for BCH(127,106) and of Majority-Logic decoding for the RM(128,99) code are both $3$ bit flips, \cite{lin2004error}. As GRAND-MO is an optimal maximum likelihood decoder, in this setting it marginally outperforms both well-known decoders by going slightly further and correctly decoding some transmissions that experienced $4$ bit flips.

When channels have memory, however, as seen for the BCH(127,106) code in Fig. \ref{fig:bch_bsc_markov_multi_burst}, the B-M decoder's performance degrades rapidly as the channel becomes more bursty. As B-M expects a BSC, that is not surprising. In contrast, GRAND-MO, which is adapted to the channel, instead gains substantially from additional structure of bursty noise. 
\begin{figure}[htbp]
\centerline{\includegraphics[width=0.49\textwidth]{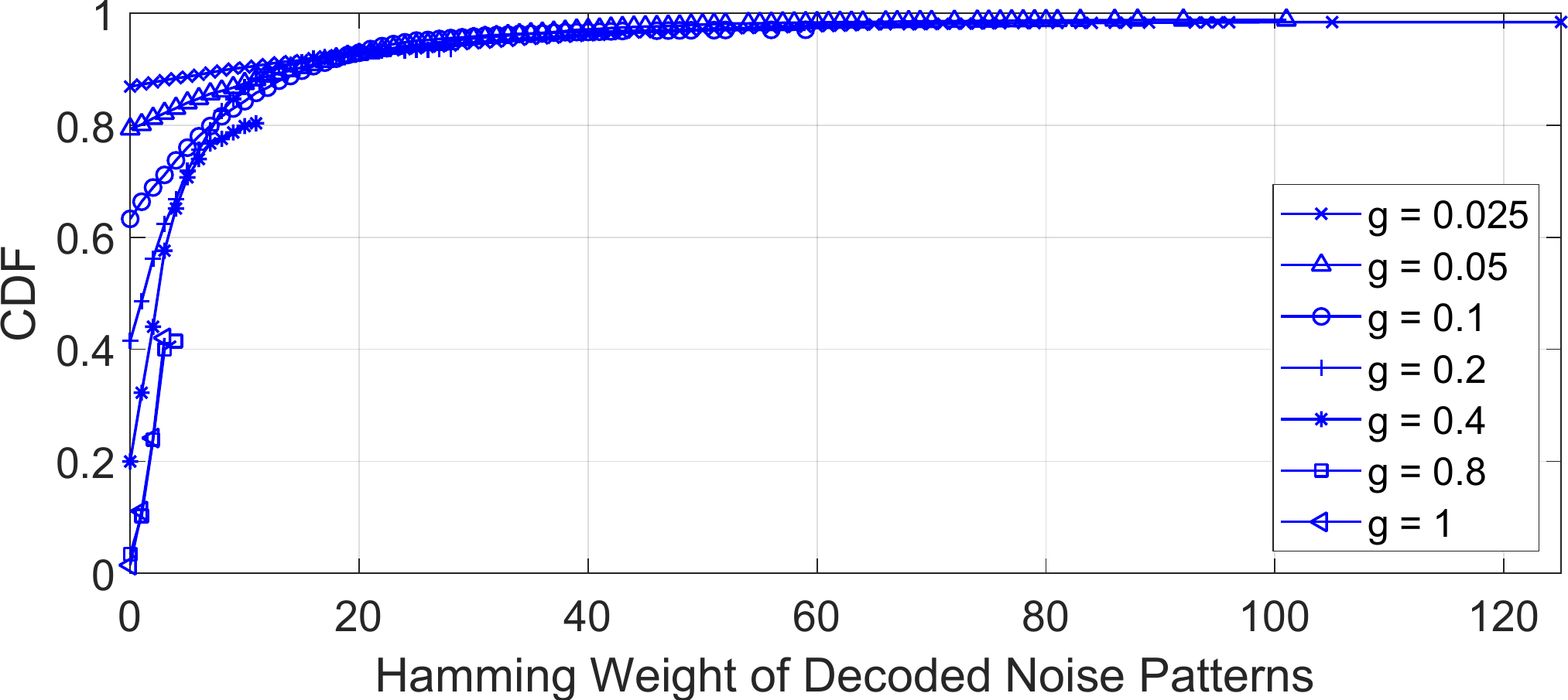}}
\caption{Empirical CDF of Hamming weights of noise patterns of successful GRAND-MO decoding of a BCH(127,106) code in a Markov channel.}
\label{fig:cdf_noise_ham}
\end{figure}
GRAND-MO's performance improvement is the result of its capability of decoding noise patterns far beyond the random-error-correcting capability of standard decoders. To illustrate this effect, we provide the empirical cumulative distribution function of Hamming weights of noise patterns that resulted in successful decodings in Fig. \ref{fig:cdf_noise_ham} for a single value of $E_b/N_0$. While the $l_{last}=3$, in bursty channels, GRAND-MO regularly consistently corrects transmission errors with more than many bit flips, and occasionally more that $100$. In bursty channels, intuition from the BSC does not carry over.

\begin{figure}[htbp]
\centerline{\includegraphics[width=0.49\textwidth]{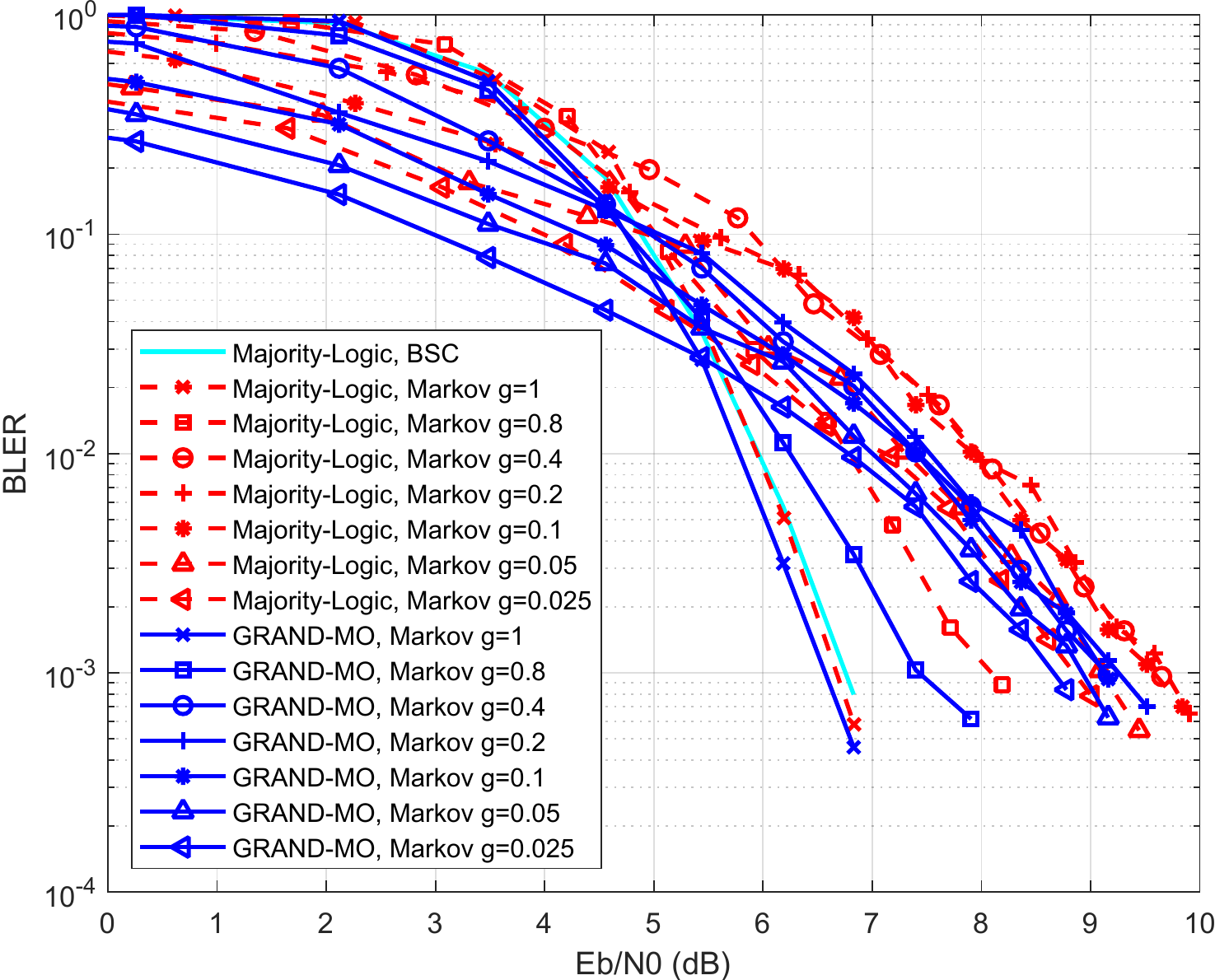}}
\caption{Performance of RM(128,99) with Majority-Logic decoder in BSC and Markov channels, and with GRAND-MO decoder in Markov channels}
\label{fig:rm_bsc_markov_multi_burst}
\end{figure}

In RM codes, code words contain groups of 1s akin to bursts in noise. One expects this mean that RM codes are not robust in bursty channels as adding a putative noise sequence that is as a burst of 1s may map to an existing, but incorrect, codeword. Fig. \ref{fig:rm_bsc_markov_multi_burst}, illustrates the performance of both GRAND-MO and Majority-Logic decoders for RM(128,99) in Markov channels.  A slight decease of value of $g$ from $1$ to $0.8$ results in more than 1.5dB loss in Majority-Logic decoder at a target BLER of $10^{-3}$. GRAND-MO performs slightly better, but still incurs almost a 1dB loss. With more memory in channels, corresponding to lower values of $g$, both decoders suffer performance degradation. This demonstrates that codes that are well-structured for BSC channels are not equal when used in bursty channels, even with an optimal decoder that is matched to the channel conditions. 

\subsection{Performance of Random Linear Codes}
\label{subsect:RLCs}

The above arguments suggest that the structure of BCH codes, which have fairly even distribution of 1s, is able to benefit from decoding that takes into account channel memory but that the structure of RM codes, which exhibit groupings of 1s, is ill-suited to such decoding. In effect, RM codes are highly tailored to the BSC assumption, whereas BCH codes, which bear some resemblance in their distribution to random codes, are well suited to  Markov channels and can be taken advantage of by GRAND-MO. 

In this vein, we explore the use of Random Linear Code (RLCs). RLCs  are linear block codes that have long been used for proofs of capacity achievability, but have been heretofore not considered practical in terms of decodability. That is entirely understandable as one requires a universal decoder to be able to decode random codes. For GRAND-MO, RLCs are no more challenging to decode than any other linear code.

\begin{figure}[htbp]
\centerline{\includegraphics[width=0.49\textwidth]{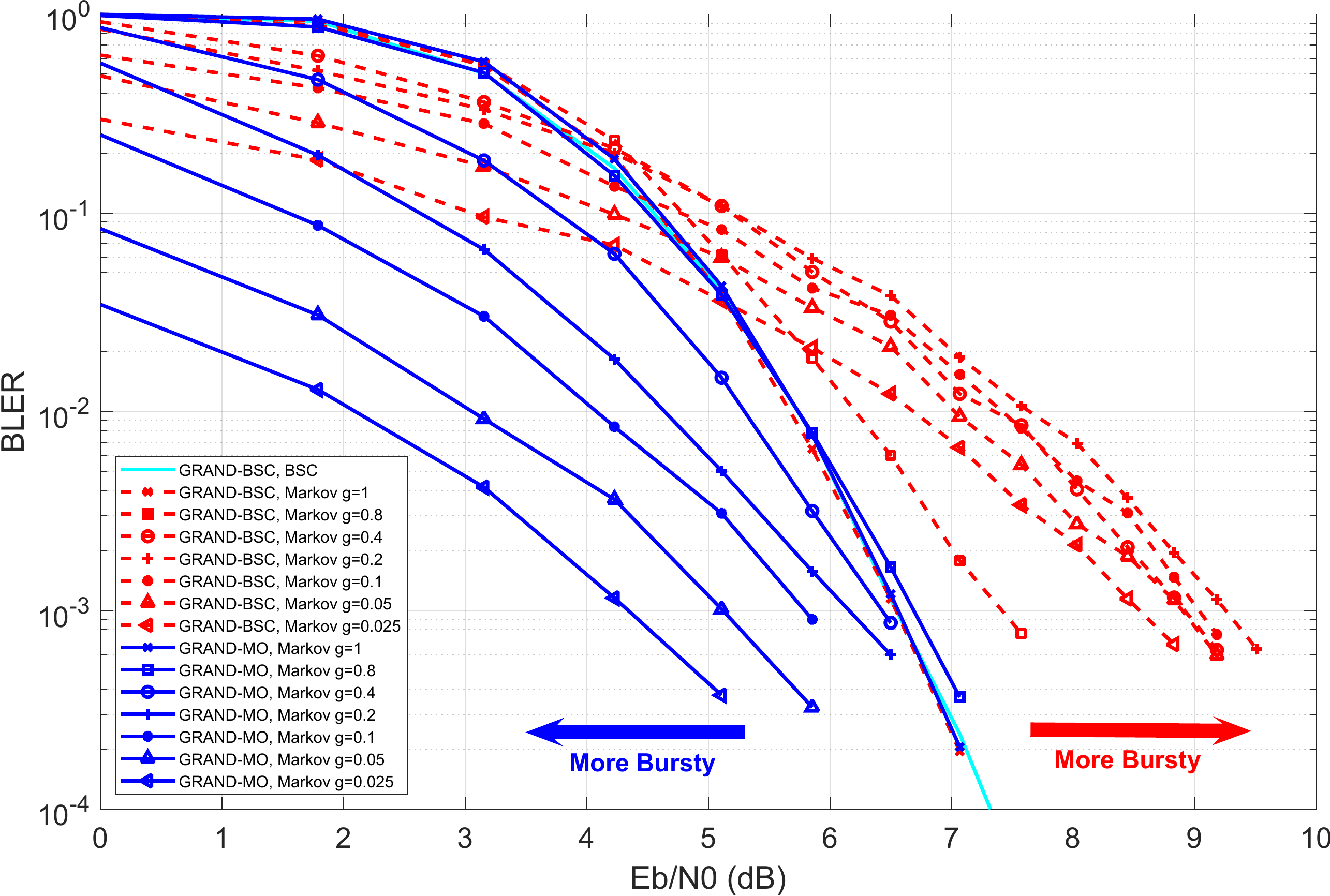}}
\caption{Performance of RLC(127,106) with GRAND-BSC (GRAND-MO with $\Delta l=0$, red lines) and channel adapted GRAND-MO (blue lines)}
\label{fig:rlc_bsc_markov_multi_burst}
\end{figure}

Unlike structured codes that only exist for certain $(n,k)$ combinations and so provide a limited set of rates, RLCs inherently provide complete flexibility of $n$ and $k$ by  adjusting dimensions of the generator matrix. This desirable property allows complete flexibility in rate-matching to channel conditions. For the results in Fig. \ref{fig:rlc_bsc_markov_multi_burst}, we randomly produced a single generator matrix creating a RLC(127,106), which enabled performance comparison with the BCH(127,106) code used for Fig. \ref{fig:bch_bsc_markov_multi_burst}. Unlike the RM code, the RLC, which inherently has limited structure, proves to be robust to bursty noise. Moreover, it provides near identical error protection to the same rate BCH code, apart from when bursts are sufficiently long on average that the BCH code's performance degrades and the RLC outperforms it. This suggests that RLCs, which have been overlooked for long codes due to the computational impracticality of their decoding, may well be suited for URLLC.

\subsection{Latency of Interleavers} \label{subsect:intl}

Interleavers effectively construct long code-words by grouping shorter code-words, resulting in significant, undesirable additional latency. Moreover, by seeking to emulate the behavior of a BSC channel, using interleavers precludes the decoding benefits, shown in Section \ref{sect:app_sim}, that taking into account memory can provide. Here we investigate the extent of interleaving required to make a Markov-channel look sufficiently BSC-like so that a standard decoder can perform acceptably and evaluate the cost, in terms of loss of $E_b/N_0$, that comes with it.

In theory, interleaving is best achieved by a random permutation, but, in practice, matrix interleavers are most commonly employed. By that method, a square matrix is created where stored coded bits awaiting transmission are written into the matrix as columns and then read out by rows. The result is that bits that are close to each other in the original order are well separated before transmission. The de-interleaver awaits receipt of all interleaved data, and then undoes the permutation before passing the correctly ordered bits to the decoder.

\begin{figure}[htbp]
\centerline{\includegraphics[width=0.49\textwidth]{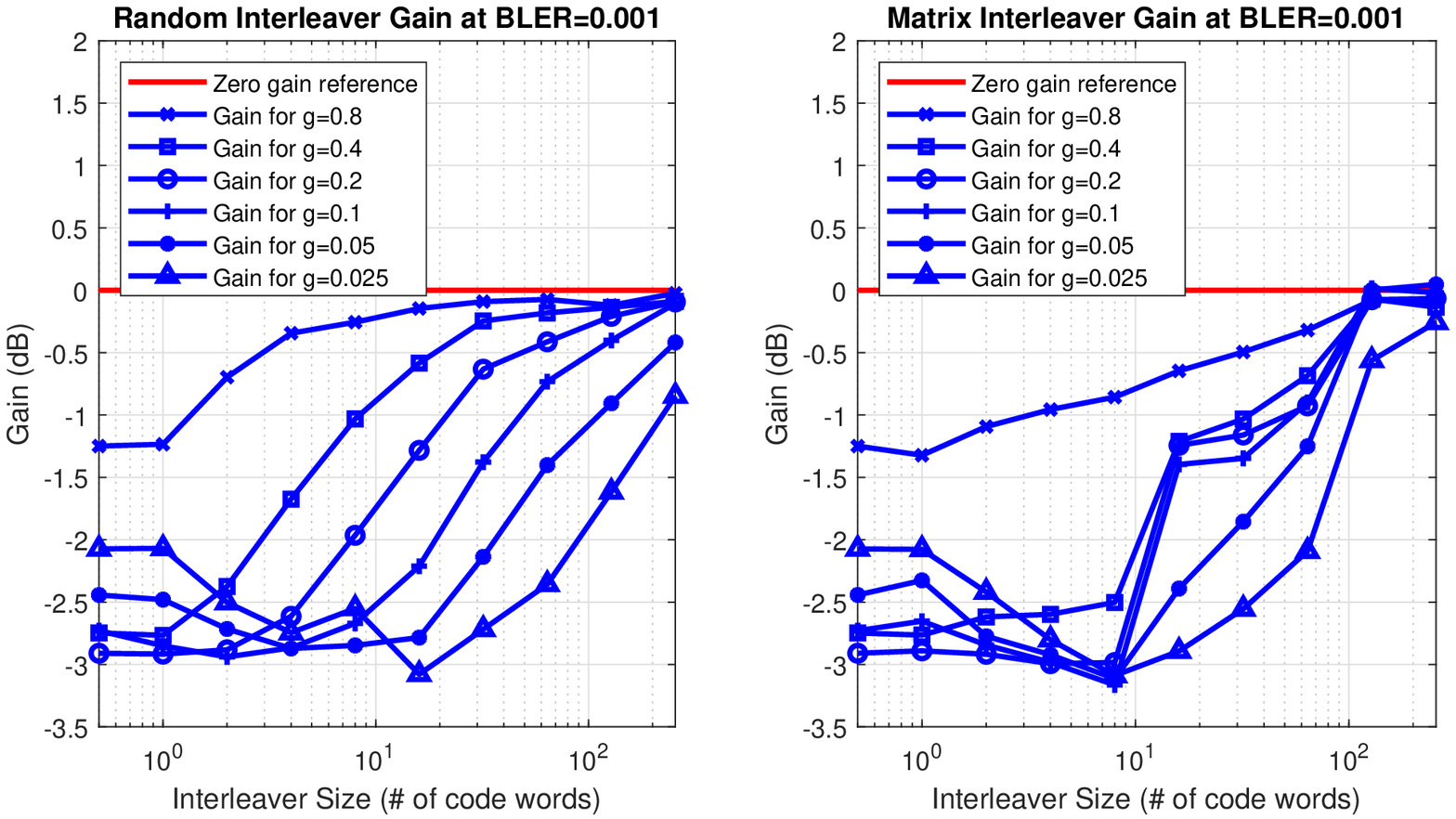}}
\caption{Performance of B-M decoders with random and matrix interleavers in Markov channels relative to the BSC and un-interleaved GRAND-MO decoding. Plotted is $E_b/N_0$ at $\text{BLER}=10^{-3}$ for a BCH(127,106).}
\label{fig:comb_comp}
\end{figure}

As a function of the number of BCH(127,106) coded packets collected by the interleaver, Fig. \ref{fig:comb_comp} presents performance in term of the loss of $E_b/N_0$ experienced by the B-M decoder in the Markov channel at a BLER of $10^{-3}$ with respect to the corresponding BSC. The gain of GRAND-MO's channel-matching performance is presented under the same condition. By not interleaving, GRAND-MO sees a consistent 1-4dB gain over the highly interleaved B-M decoder. Moreover, to obtain that BSC performance it is necessary to interleave over hundreds of packets, amounting to tens of thousands of bits, which necessitates substantial unwanted delays. By ditching interleavers and using GRAND-MO, we gain both improved BLER performance and have eliminated an unwanted source of delay in the process.

\section{Summary} \label{sect:conclusion}

The drive to enable ultra reliable low latency communications for modern applications is hindered by the assumption common to standard decoders that channels are memoryless. To match that assumption, interleavers buffer large volumes of data before transmission and deinterleavers must wait for all of that data to be delivered before decoding can begin, resulting in unwanted delays. To circumvent interleavers  requires decoders that can directly manage channels subject to burst errors.

As one potential solution, here we further develop the Guessing Random Additive Noise Decoding (GRAND) methodology to make it suitable for use with channels experiencing correlated errors. By analysing a Markov model of channel noise, we develop an efficient algorithm for sequentially producing putative noise patterns in order of their likelihood. When used in GRAND, it results in GRAND-MO, a hard detection decoder that can  accurately decode any short, high-rate code. Using GRAND-MO we establish that not only are their latency reductions to be had by ditching interleavers, but there are also significant BLER performance gains to be availed of by the retention and use of statistical structure within the noise.

Using GRAND-MO's ability to decode any code, we find that performance with structured Reed-Muller and BCH codes ultimately degrades as channels become more bursty. Random Linear Codes, which have been little investigated outside of theory, however, appear to be robust to channel conditions. As a result, one possible way forward is to ditch interleavers, and use RLCs decoded by GRAND-MO to enable accurate, high-rate ultra-low latency communications.

\bibliographystyle{IEEEtran}
\bibliography{my_bibli}

\end{document}